\documentstyle [12pt] {article}

\newcommand{\beq}{\begin{equation}}
\newcommand{\eeq}{\end{equation}}
\newcommand{\beqs}{\begin{eqnarray}}
\newcommand{\eeqs}{\end{eqnarray}}

\newcommand{\Z}{\mbox{\rm Z$\!\!$Z}}

\catcode`@=11
\@addtoreset{equation}{section}
\@addtoreset{equation}{subsection}
\def\theequation{\ifnum\value{section}=0 \arabic{equation}\ignorespaces
\else \ifnum\value{section}=-1 A.\arabic{equation}\ignorespaces
\else \ifnum\value{subsection}=0 \thesection.\arabic{equation}\ignorespaces
\else \thesection.\arabic{subsection}.\arabic{equation}\ignorespaces
                           \fi
                      \fi
                 \fi}
\catcode`@=12

\begin{document}

\def\thefootnote{\fnsymbol{footnote}}

\baselineskip 6.0mm

\begin{flushright}
ITP-SB-97-30
\end{flushright}

\vspace{4mm}

\begin{center}

{\Large \bf A Mapping Relating Complex and Physical Temperatures in 
the 2D $q$-State Potts Model and Some Applications} 

\vspace{8mm}

\setcounter{footnote}{0}
Heiko Feldmann\footnote{email: feldmann@insti.physics.sunysb.edu},
\setcounter{footnote}{6}
Robert Shrock\footnote{email: shrock@insti.physics.sunysb.edu}
\setcounter{footnote}{7}
and Shan-Ho Tsai\footnote{email: tsai@insti.physics.sunysb.edu}

\vspace{6mm}

Institute for Theoretical Physics  \\
State University of New York       \\
Stony Brook, N. Y. 11794-3840  \\

\vspace{10mm}

{\bf Abstract}
\end{center}

     We show an exact equivalence of the free energy of the $q$-state 
Potts antiferromagnet on a lattice $\Lambda$ for the full temperature 
interval $0 \le T \le \infty$ and the free energy of the $q$-state Potts 
model on the dual lattice for a semi-infinite interval of complex 
temperatures (CT). This implies the existence of two quite different types 
of CT singularities: the generic kind, which does not obey universality or 
various scaling relations, and a special kind which does obey such properties 
and encodes information of direct physical relevance.  We apply this 
observation to characterize CT properties of the Potts model on several 
lattices, to rule out two existing conjectures, and to determine the critical 
value of $q$ above which the Potts antiferromagnet on the diced lattice has 
no phase transition. 

\vspace{16mm}

\pagestyle{empty}
\newpage

\pagestyle{plain}
\pagenumbering{arabic}
\renewcommand{\thefootnote}{\arabic{footnote}}
\setcounter{footnote}{0}

   The study of statistical mechanical models with magnetic field \cite{yl},
temperature \cite{mef}-\cite{dg}, or both \cite{ih} generalized from 
real to complex values has yielded interesting insights into the properties 
of these models.  Complex-temperature (CT) singularities have been shown to 
have, in general, rather different properties than physical critical points of 
spin models, including violation of exponent relations such as 
$\gamma \ne \gamma'$ \cite{ms} and 
violation of universality, as evidenced by lattice dependence of exponents 
(shown using exact results) \cite{chisq}.  It is therefore of fundamental 
interest to understand better how CT properties of spin models are related to 
physical properties.  Here we shall show an exact equivalence of the free 
energy of the $q$-state 
Potts antiferromagnet on a lattice $\Lambda$ for the full temperature
interval $0 \le T \le \infty$ and the free energy of the $q$-state Potts
model ($q$PM) on the dual lattice for the interval
 $-\infty \le \exp(J/(k_BT)) \le 
-(q-1)$, where $J$ denotes the spin-spin coupling. 
This result leads to the important conclusion 
that there are two different types of CT singularities: the generic type, with 
violations of scaling and universality, and a special type, which is 
closely related to physical critical points and can be described 
by the same ideas of scaling, renormalization group (RG), universality classes
determined as RG fixed points and related, for 2D models, to 
conformal field theories.  Our observation also yields a way to determine some 
properties of certain theories with non-Gibbs measures, by relating them 
exactly to theories with Gibbs measures.  This is a useful connection, since 
theories with non-Gibbs measures appear not just in the context of complex 
temperature or magnetic field, but also in a number of physical situations, 
such as quantum spin models in condensed matter, lattice quantum 
chromodynamics with finite chemical potential or topological charge, and the 
lattice formulation of the standard SU(2) $\times$ U(1) electroweak gauge 
theory. 

   The $q$-state Potts model \cite{potts}-\cite{mbook} with zero external 
field(s) on a lattice $\Lambda$ is defined by the 
partition function $Z = \sum_{ \{ \sigma_n \} } \exp(-\beta {\cal H})$ with 
the Hamiltonian
\beq
{\cal H} = -J \sum_{\langle nn' \rangle} \delta_{\sigma_n \sigma_{n'}}
\label{hpotts}
\eeq
in standard notation, where $\sigma_n=1,...,q$, $\beta = (k_BT)^{-1}$, and 
$\langle n n' \rangle$ denotes pairs of nearest-neighbor sites on $\Lambda$.  
We also define $K = \beta J$, 
\beq
a = z^{-1} = e^{K} \ , \quad x = \frac{e^K-1}{\sqrt{q}}
\label{azx}
\eeq
We consider $d=2$ dimensional lattices, since in this case the model with
spin-spin interactions along bonds maps to another with similar interactions on
the bonds of the dual lattice.  Physical applications of the 2D 
ferromagnetic (FM) Potts model for $q=3$ and 4 include
modelling properties of monolayers of gas molecules absorbed on substrates.  
The 2D Potts ferromagnet is known to have a ($\Z_q$ symmetry-breaking) 
phase transition which is second-order for $2 \le q \le 4$ and first order for
$q \ge 5$.  The critical exponents and universality classes of the cases where
the model has second-order transitions are well understood \cite{wurev,cft}, 
but, aside from the $q=2$ Ising case \cite{ons}, the free energy has never 
been calculated for general $T$. The Potts antiferromagnet has also been of 
interest because of its connection with graph colorings and the fact that, 
for certain lattices and values of $q$, it exhibits nonzero ground state 
entropy. 

   Now consider the $q$PM on a 2D lattice $\Lambda$ with 
$N_0$ sites (0-cells), $N_1$ bonds (1-cells) and $N_2$ faces (2-cells). 
The dual lattice $\Lambda_d = {\cal D}(\Lambda)$ is defined by associating 
uniquely a $(2-p)$-cell of $\Lambda_d$ with each $p$-cell of $\Lambda$. 
The partition function of the $q$PM satisfies the duality relation 
\cite{potts,wurev,pdual}
\beq
Z(\Lambda,q,a) = x^{N_1}q^{N_0-1-(1/2)N_1}Z(\Lambda_d,q,a_d)
\label{zduality}
\eeq
where 
\beq
a_d \equiv {\cal D}(a) = \frac{a+q-1}{a-1} \ , \quad i.e.\quad 
x_d = \frac{1}{x}
\label{ad}
\eeq
It follows from (\ref{zduality}) that in the thermodynamic limit, the
singularities of the free energy of the $q$PM on $\Lambda$ at a point 
$a$ satisfy
\beq
f(\Lambda,q,a)_{sing} = f(\Lambda_d,q,a_d)_{sing}
\label{fduality}
\eeq
The key
observation is that under duality, the complete physical temperature interval 
$0 \le T \le \infty$, i.e., $0 \le a \le 1$ of the $q$-state Potts
antiferromagnet (qPAF) on $\Lambda$ is mapped to the CT interval
$-\infty \le a_d \le -(q-1)$ on $\Lambda_d$ and vice versa. (In contrast, the
physical $T$ interval of the $q$-state Potts ferromagnet on $\Lambda$,
$1 \le a \le \infty$, maps under duality to another physical $T$ interval, 
$1 \le a_d \le \infty$, of the same model on $\Lambda_d$.) 

   From the above observation we can derive several interesting results.
First, from eq. (\ref{fduality}), it follows that there is an exact 1-1
correspondence between the CT singularities of the $q$PM on $\Lambda_d$ in the
interval $-\infty \le a_d \le -(q-1)$ and the physical singularities of the
$q$-state Potts AF on $\Lambda$.  In particular, if the $q$PAF on $\Lambda$ 
has a phase transition at $a_{PM-AFM}$ (here, PM = paramagnetic), then there 
is a corresponding CT singularity in $f(\Lambda_d,q,a)$ at $a=a_\ell$, where
\beq
a_\ell={\cal D}(a_{PM-AFM})
\label{aell}
\eeq
and this is the
leftmost CT singularity in the interval $-\infty < a_d \le 0$ (hence the label
$\ell$).  These transitions are of the same order.  If the PM-AFM transition of
the $q$PAF on $\Lambda$ is continuous, then the singularity in $f$ of the 
$q$PM 
at the CT point $a_\ell$ is the same when approached from the left or from the
right and, writing $f(\Lambda_d,q,a)_{sing} \sim
|a-a_\ell|^{2-\alpha_\ell}$, we have $\alpha_\ell=\alpha=\alpha'$, where the 
exponent $\alpha=\alpha'$ is the physical specific heat exponent of the 
$q$-state PAF at the PM-AFM transition. 
(In general we cannot use duality to relate the $\beta$
or $\gamma$ exponents for the staggered magnetization and susceptibility of the
$q$-state PAF at $a_{PM-AFM}$ on $\Lambda$ and staggered or uniform $\beta$ 
or $\gamma$ exponents at $a_\ell$ since the duality applies in the absence of 
external field(s).)   Now for a model above its lower critical
dimensionality, so that there is a symmetry-breaking phase transition, a
general property of the CT phase diagram is that in the complex $a$ plane, 
the continuous locus of points ${\cal B}$ where the free energy is 
nonanalytic (which determines the CT phase boundaries) is compact, 
i.e., the region sufficiently far from the origin is in the 
complex-temperature extension of the FM phase \cite{ih,chisq}.  (Henceforth, we
shall use (CTE)FM to denote the CT extension of the FM phase, and so forth 
for other phases.) It follows that if the $q$PAF on $\Lambda$ has a PM-AFM
transition, then since $a_\ell$ is the leftmost singularity in the $a_d$ 
plane of the $q$-state Potts model on $\Lambda_d$, the CT phase to the left 
of $a_\ell$ is the (CTE)FM phase of the latter model.  

   It may be helpful to illustrate these general results briefly for the known 
$q=2$ case. For the square lattice, $\Lambda=\Lambda_d$ and 
$a_{sq,q=2,PM-AFM}=\sqrt{2}-1$, so that $a_{sq,q=2,\ell}=-(\sqrt{2}+1)$. In the
terminology of Ref. \cite{chisq}, the CT phase to the right of 
$a_{sq,q=2,\ell}$ is an O phase, meaning that it is not the CT extension of 
any physical phase.  For the honeycomb (hc) lattice, 
$a_{hc,q=2,PM-AFM}=2-\sqrt{3}$, so that the leftmost CT singularity of the 
Ising model on the dual, triangular (t) lattice is at 
$a_{t,q=2,\ell}=-\sqrt{3}$.  For the Ising model on the honeycomb lattice, 
the phases on the left and right of $a_{hc,q=2,\ell}$ are the (CTE)FM and 
(CTE)AFM phases; for the Ising model on the triangular lattice, the phases to
the left and right of $a_{t,q=2,\ell}$ are the (CTE)FM and an O phase,
respectively \cite{chitri}.  The analyticity of $f$ on the hc lattice in the 
interval $-\infty < a_d < -1$ is now seen as being equivalent to the absence 
of a finite-$T$ transition in the Ising AF on the triangular lattice.  
On the kagom\'e lattice,
$a_{kag,q=2,\ell} = -3^{1/4}(2-\sqrt{3})^{-1/2}$ \cite{kag,cmo}, separating 
the (CTE)FM and (CTE)PM phases on the left and right, and corresponding to the
PM-AFM transition of the Ising AF on the dual diced lattice. 

A second type of behavior occurs if the $q$PAF on $\Lambda$ has no finite-$T$ 
phase transition but is critical at $T=0$; this is associated with a part of 
the CT phase boundary ${\cal B}$ passing through $a=0$ in the CT phase
diagram of this model.  It then follows that for the $q$-state Potts model on 
$\Lambda_d$ a part of the respective ${\cal B}$ for that model passes through 
$a_d=-(q-1)$.  As examples, the Ising AF on the triangular lattice, and the 
$q=3$ Potts AF on the square and kagom\'e lattices all have 
zero-temperature critical points, so that there are respective singular points
at the dual images, $a_{hc,q=2,\ell}=-1$, $a_{sq,q=3,\ell}=-2$, and 
$a_{diced,q=3,\ell}=-2$ on the dual lattices.   As discussed for the square 
lattice in Ref. \cite{pfef}, $a_{sq,q=3,\ell}=-2$ is the leftmost point where 
the CT phase boundary ${\cal B}$ crosses the negative real axis (see also
Refs. \cite{mbook,wuz}). 

    As the third type of behavior, if the $q$-state Potts AF on $\Lambda$ 
has no finite-temperature phase transition and is also
not critical at $T=0$ (which is manifested by ${\cal B}$ not passing through
$a=0$ in the complex $a$ plane), then it follows that $f(\Lambda_d,q,a_d)$ is 
analytic in the CT interval $-\infty < a_d \le -(q-1)$, which therefore must 
be part of the (CTE)FM phase of the $q$PM on $\Lambda_d$.  This type of 
behavior occurs, for example, for the Potts model on the square lattice with 
$q \ge 4$. 

    We consider now the $q=3$ Potts model on the honeycomb and triangular 
lattices.  Usually, in cases where there are no exact solutions, the locations
of CT singular points are most accurately determined from series analyses; 
indeed, the complications due to these singularities were recognized in early 
work on series \cite{dg}.  However, for $a_\ell$, one has a powerful
alternative approach: to locate it using eq. (\ref{aell}) and a precise
determination of the physical PM-AFM transition of the $q$PAF on the dual
lattice.  This is our first application.  For the $q=3$ PAF on the triangular 
lattice, Monte Carlo and series analyses \cite{grest}-\cite{adler} led to
the conclusion that this model has a weakly first order transition.  The most
recent study \cite{adler}, using Monte Carlo methods, has yielded a very 
precise determination of the transition point: 
$T_t = 0.62731 \pm 0.00006$, i.e., $a_{PM-AFM,t} = 0.20309 \pm
0.00003$. We infer that the $q=3$ Potts model on the honeycomb lattice has 
a corresponding singularity at the CT point 
\beq
a_{hc,q=3,\ell}={\cal D}(a_{PM-AFM,t})=-(2.76454 \pm 0.00015)
\label{aellhcq3}
\eeq
i.e., $z_{hc,q=3,\ell}=a_{hc,q=3,\ell}^{-1}=-(0.36172 \pm 0.00002)$.  This 
value is in agreement with, and is more accurate than, the value 
$z_{hc,q=3,\ell}=-(0.363 \pm 0.003)$ obtained from a recent low-temperature 
series analysis in Ref. \cite{jge} (where $z_{hc,q=3,\ell}$ is denoted 
$u_-$).  If one accepts the reported result \cite{grest}-\cite{adler} that 
the PM-AFM transition of the 
$q=3$ PAF on the triangular lattice is (weakly) first order, the same applies 
for the singularity at $a_{hc,q=3,\ell}$, and, furthermore, these transitions 
have the same value of the latent heat.  If one were formally to assign an
exponent $\alpha$ to these transitions, it would thus be $\alpha=1$.  Past
experience shows that it can be difficult to distinguish a weakly first order
transition from a continuous one, and indeed the series analysis in Ref. 
\cite{jge} found evidence for a continuous transition at $z_{hc,q=3,\ell}$ 
with $\alpha \simeq 0.5$.  

    A second application is to test a recent conjecture for an exact 
analytic value of the CT singularity at $z_{hc,q=3,\ell}$ made in Ref. 
\cite{m95}, viz., that 
$z_{hc,q=3,\ell}=\cos(2\pi/9)-3^{1/2}\sin(2\pi/9)=-.347296355..$, i.e., 
$a_{hc,q=3,\ell} = -2.87938524...$ (arising as a root of the equation 
$z^3-3z-1=0$). 
Since the result of Ref. \cite{jge} was $5\sigma$ (where $\sigma=0.003$ was the
uncertainty in the location of the singular point) away from this conjecture, 
it was concluded that this conjecture is unlikely but not impossible to be 
correct.  We can strengthen this conclusion here: our determination of 
$z_{hc,q=3,\ell}$ decisively refutes the conjecture of Ref. \cite{m95}. 

   As a third application, we use our recent study of the $q=3$ Potts AF on the
honeycomb lattice \cite{p3afhc}, where we found that this model has no
finite-temperature PM-AFM transition and is, indeed, disordered at $T=0$, with
nonzero ground state entropy $S_0/k_B=0.5068(3)$.  From our discussion above,
this implies that the free energy of the $q=3$ Potts model on the triangular 
lattice is analytic in the interval $-\infty < a \le -2$.  In passing, we note
that this rules out yet another conjecture \cite{mm}, that there could be
a singularity in this model at the largest negative root of the equation 
$a^3+6a^2+3a-1=0$, viz., $a=-5.411474...$ (associated with the completion of
complex-conjugate branches of CT zeros labelled 6 in Ref. \cite{mm}). Our
result also implies that for the $q=3$ Potts model on the triangular lattice 
the leftmost component of the CT phase boundary ${\cal B}$ that crosses the 
real $a$ axis must do so at $a > -2$.  These results apply 
{\it a fortiori} to the case $q \ge 4$: in this range, the Potts
AF on the honeycomb lattice has no finite-$T$ phase transition and is
disordered, with finite entropy, at $T=0$, so that for each respective value of
$q$, the corresponding Potts model on the triangular lattice is analytic in the
range $-\infty < a \le -(q-1)$.   These results are borne out by explicit
calculations of CT zeros of the partition function for $q$-state Potts models
on honeycomb and kagom\'e lattices, to be presented elsewhere.  

   A fourth application concerns the $q$-state PAF on the diced lattice (dual
of kagom\'e).  Ref. \cite{jge} found CT singularities at 
$z_{kag,q=3,\ell}=-0.4023 \pm 0.0005$ and $z_{kag,q=4,\ell}=-0.42 \pm 0.01$ 
for the $q=3$ and $q=4$ Potts model on the kagom\'e lattice.  From the first of
these, we calculate that the $q=3$ PAF on the diced lattice has a
PM-AFM phase transition at
\beq
a_{diced,q=3,PM-AFM} = 0.1393 \pm 0.0008
\label{adicedq3}
\eeq
For comparison, from the exact value for $a_{kag,q=2,\ell}$, we have 
$a_{diced,q=2,PM-AFM}=0.4354205...$ for the Ising case.  Furthermore, we find 
the interesting result that the $q=4$ Potts AF on the diced lattice has no 
finite-$T$ phase  transition and also is not critical at $T=0$, since
\beq
{\cal D}(a_{kag,q=4,\ell})=-(0.18 \pm 0.02) 
\label{adicedq4}
\eeq
is negative.  Remarkably, we have thus used information on a CT singular
point of the Potts model on one lattice to derive a new physical result for
this model on the dual lattice.  From this we also determine the critical
integer value of $q$ for the $q$-state Potts antiferromagnet on the diced 
lattice above which it has no finite-temperature phase transition, viz., 
$q=3$. 

   In summary, we have used an exact duality property of a 2D $q$-state Potts 
model to relate certain complex-temperature properties of the model on a 
given lattice to physical properties of the Potts antiferromagnet on the dual
lattice.  This result gives useful information about the complex-temperature
phase diagrams of 2D Potts models, and we have been able to use it to rule out 
two conjectures and to find the critical integer value of $q$ for the Potts 
antiferromagnet on the diced lattice. 

This research was supported in part by the NSF grant PHY-93-09888.  R.S. thanks
Prof. Tony Guttmann for giving us a copy of Ref. \cite{jge} prior to
publication.

\vfill
\eject

\vspace{6mm}

\begin{thebibliography}{99}

\bibitem{yl}{Yang, C. N. and Lee, T. D. 1952 {\it Phys. Rev.} {\bf 87} 404;
Lee, T. D. and Yang, C. N. 1952 {\it ibid} {\bf 87} 410.}

\bibitem{mef}{Fisher, M. E. 1965 {\it Lectures in Theoretical Physics}
(Univ. of Colorado Press), vol. 7C, p. 1.} 

\bibitem{kat}{Katsura, S. 1967 {\it Prog. Theor. Phys.} {\bf 38}, 1415;
Abe, R. 1967 {\it Prog. Theor. Phys.} {\bf 38}, 322;
Ono, S., Karaki, Y., Suzuki, M., and Kawabata, C. 1968 {\it J. Phys. Soc.
Jpn.} {\bf 25}, 54.}

\bibitem{dg}{Domb, C. and Guttmann, A. J. 1970 {\it J. Phys.} C {\bf 3} 1652.}

\bibitem{ih}{Matveev, V. and Shrock, R. 1995 {\it J. Phys. A} {\bf 28} 4859; 
Matveev, V. and Shrock, R. 1996 {\it Phys. Rev.} {\bf E53}, 254; 1996
{\it Phys. Lett.} {\bf A215} 271.} 

\bibitem{ms}{Marchesini, G. and Shrock, R. 1989. {\it Nucl. Phys.} 
{\bf B318} 541.}

\bibitem{chisq}{Matveev, V. and Shrock, R. 1995 {\it J. Phys. A} {\bf 28} 
1557.}

\bibitem{potts}{Potts, R. B. 1952 {\it Proc. Camb. Phil. Soc.} {\bf 48} 
106.}

\bibitem{wurev}{Wu, F. Y. 1982 {\it Rev. Mod. Phys.} {\bf 54} 235.} 

\bibitem{mbook}{Martin, P. P. 1991 {\it Potts Models and Related Problems in
Statistical Mechanics} (World Scientific, Singapore).}

\bibitem{cft}{Itzykson, C., Saleur, H., and Zuber, J.-B., eds. 1988
{\it Conformal Invariance and Applications to Statistical
Mechanics} (World Scientific, Singapore).}

\bibitem{ons}{Onsager, L. 1944 {\it Phys. Rev.} {\bf 65} 117; 
Yang, C. N. 1952 {\it Phys. Rev.} {\bf 85} 808.}

\bibitem{pdual}{Kim, D. and Joseph, R. J. 1974 {\it J. Phys.} C {\bf 7}, 
L167; Burkhardt, T. W. and Southern, B. W. 1978 {\it J. Phys. A} {\bf 11} 
L247.}

\bibitem{chitri}{Matveev, V. and Shrock, R. 1996 {\it J. Phys. A} {\bf 29} 
803.}

\bibitem{kag}{Abe, R., Dotera, T., and Ogawa, T. 1991 {\it Prog. Theor. Phys.} 
{\bf 85} 509.} 

\bibitem{cmo}{Matveev, V. and Shrock, R. 1995 {\it J. Phys. A} {\bf 28} 5235.}

\bibitem{pfef}{Matveev, V. and Shrock, R. 1996 {\it Phys. Rev.} {\bf E54} 
6174.}

\bibitem{wuz}{Chen, C. N., Hu, C. K, and Wu, F. Y. 1996 {\it Phys. Rev. Lett.}
{\bf 76} 169.} 

\bibitem{grest}{Grest, G. S. 1981 {\it J. Phys. A} {\bf 14} L217;
Saito, Y. 1982 {\it J. Phys. A} {\bf 15} 1885.}

\bibitem{entingwu}{Enting, I. G. and Wu, F. Y. 1982 {\it J. Stat. Phys.} 
{\bf 28} 351.}

\bibitem{adler}{Adler, J., Brandt, A., Janke, W., and Shmulyian, S. 1995
{\it J. Phys. A} {\bf 28}, 5117.}

\bibitem{jge}{Jensen, I., Guttmann, A. J., and Enting, I. G. 1997
``The Potts Model on Kagom\'e and Honeycomb Lattices'', J. Phys. A, to appear.}

\bibitem{m95}{Maillard, J.-M. {\it Computers and Mathematics with
Applications}, to appear.}

\bibitem{p3afhc}{Shrock, R. and Tsai, S.-H. 1997 {\it J. Phys. A} {\bf 30} 
495; {\it Phys. Rev.} {\bf E55} 5165; {\it ibid} 6791.}

\bibitem{mm}{Martin, P. P. and Maillard, J.-M. 1986 {\it J. Phys. A} {\bf 19} 
L547.}

\end{thebibliography}
\end{document}